\documentclass[12pt]{article}
\usepackage{axodraw,bbold}

\catcode`@=12
\begin{document}
\thispagestyle{empty}
\begin{flushright} 
UCRHEP-T429\\ 
July 2007\
\end{flushright}
\vspace{0.5in}
\begin{center}
{\LARGE	\bf New Lepton Family Symmetry and\\ 
Neutrino Tribimaximal Mixing\\}
\vspace{1.5in}
{\bf Ernest Ma\\}
\vspace{0.2in}
{\sl Physics and Astronomy Department\\
University of California, Riverside\\
Riverside, California 92521, USA\\}
\vspace{1.5in}
\end{center}

\begin{abstract}\
The newly proposed finite symmetry $\Sigma(81)$ is applied to the problem 
of neutrino tribimaximal mixing.  The result is more satisfactory than those 
of previous models based on $A_4$ in that the use of auxiliary symmetries 
(or mechanisms) may be avoided.  Deviations from the tribimaximal pattern 
are expected, but because of its basic structure, only $\tan^2 \theta_{12}$ 
may differ significantly from 0.5 (say 0.45) with $\sin^2 2 \theta_{23}$ 
remaining very close to one, and $\theta_{13}$ very nearly zero.
\end{abstract}

\newpage
\baselineskip 24pt

Based on present neutrino-oscillation data, the neutrino mixing matrix 
$U_{\alpha i}$ linking the charged leptons ($\alpha = e, \mu, \tau$) to 
the neutrino mass eigenstates ($i = 1, 2, 3$) is determined to a large 
extent \cite{s06}.  In particular, a good approximate description is 
that of the so-called tribimaximal mixing of Harrison, Perkins, and Scott 
\cite{hps02}, i.e.
\begin{equation}
U_{\alpha i} = \pmatrix{ \sqrt{2/3} & \sqrt{1/3} & 0 \cr -\sqrt{1/6} & 
\sqrt{1/3} & -\sqrt{1/2} \cr -\sqrt{1/6} & \sqrt{1/3} & \sqrt{1/2}}.
\end{equation}
Using the discrete lepton family symmetry group $A_4$ \cite{mr01,bmv03}, 
this pattern has been discussed in a number of recent papers with varying 
additional assumptions \cite{m04,af05,m05,bh05,z05,af06,m06_1,hkv06,afl06,
km06,m06_4,h06}.  In particular, auxiliary symmetries (or mechanisms) beyond 
$A_4$ are required to enforce the following conflicting alignment of 
vacuum expectation values: (1,1,1) for a {\bf 3} representation which couples 
to charged leptons, and (1,0,0) for a {\bf 3} representation which 
couples to neutrinos.  As shown below, this problem may be alleviated if 
$A_4$ is replaced by another finite discrete symmetry $\Sigma(81)$, which 
was recently proposed \cite{m06_5}.

Consider the basis $(a_1,a_2,a_3)$ and the $Z_3$ transformation
\begin{equation}
a_1 \to a_2 \to a_3 \to a_1.
\end{equation}
If this is supplemented with the $Z_2$ transformation
\begin{equation}
a_{1,2} \to -a_{1,2}, ~~~ a_3 \to a_3,
\end{equation}
then the group generated is $A_4$, which is the symmetry group of the even 
permutation of 4 objects, and that of the perfect tetrahedron \cite{m02}. 
It is also a subgroup of $SU(3)$, denoted as $\Delta(12)$.  If Eq.~(2) is 
supplemented instead with another $Z_3$, i.e.
\begin{equation}
a_1 \to a_1, ~~~ a_2 \to \omega a_2, ~~~ a_3 \to \omega^2 a_3,
\end{equation}
where $\omega = \exp (2 \pi i/3) = -1/2 + i \sqrt{3}/2$, then the group 
generated is $\Delta(27)$ \cite{bgg84,mvkr06,m06_6}, which is also a 
subgroup of $SU(3)$.  If Eq.~(4) is replaced with
\begin{equation}
a_1 \to \omega a_1, ~~~ a_{2,3} \to a_{2,3}, 
\end{equation}
then the group generated, call it $\Sigma(81)$, contains $\Delta(27)$. 
It is a subgroup of $U(3)$ but not $SU(3)$.  [Note that Eqs.~(2) and (5) 
are sufficient for generating the four transformations used to obtain 
$\Sigma(81)$ in Ref.~\cite{m06_5}.] It has 9 one-dimensional 
irreducible representations ${\bf 1_i}(i=1,...,9)$ and 8 three-dimensional 
ones ${\bf 3_A,\bar{3}_A,3_B,\bar{3}_B,3_C,\bar{3}_C,3_D,\bar{3}_D}$.  Its 
$17 \times 17$ character table and the 81 matrices of its defining 
representation ${\bf 3_A}$ are given in Ref.~\cite{m06_5}.

Consider the supersymmetric extension of the Standard Model with 3 lepton 
families. Under $\Sigma(81)$, let
\begin{eqnarray}
&& L_i = (\nu_i,l_i) \sim {\bf 3_A}, ~~~ l^c_i \sim {\bf 1_{1,2,3}}, ~~~ 
\Phi = (\phi^0, \phi^-) \sim {\bf 1_1}, \\
&& \sigma_i \sim {\bf 3_A}, ~~~ \bar{\sigma}_i \sim {\bf \bar{3}_A}, ~~~ 
\chi_i \sim {\bf 3_B}, ~~~ \bar{\chi}_i \sim {\bf \bar{3}_B}, ~~~ 
\xi = (\xi^{++}, \xi^+, \xi^0) \sim {\bf 1_1}.
\end{eqnarray}
Using the multiplication rules given in the Appendix, the allowed 
quadrilinear Yukawa terms are $(L_1 \bar{\sigma}_1 + L_2 \bar{\sigma}_2 
+ L_3 \bar{\sigma}_3) l^c_1 \Phi$, $(L_1 \bar{\sigma}_1 + \omega^2 L_2 
\bar{\sigma}_2 + \omega L_3 \bar{\sigma}_3) l^c_2 \Phi$, $(L_1 \bar{\sigma}_1 
+ \omega L_2 \bar{\sigma}_2 + \omega^2 L_3 \bar{\sigma}_3) l^c_3 \Phi$, 
$(L_1 L_1 \sigma_1 + L_2 L_2 \sigma_2 + L_3 L_3 \sigma_3) \xi$, and 
$(L_1 L_2 \chi_3 + L_2 L_3 \chi_1 + L_3 L_1 \chi_2) \xi$.  As shown below, 
the singlet superfields $\sigma_i$, $\bar{\sigma}_i$, $\chi_i$, $\bar{\chi}_i$
will acquire vacuum expectation values \underline{without} breaking the 
supersymmetry.  The desirable solutions (1,1,1) for $\sigma_i$, 
$\bar{\sigma}_i$, and (1,0,0) for $\chi_i$ and $\bar{\chi}_i$ may then be 
obtained in a natural symmetry limit, for which the mismatch between the 
charged-lepton and neutrino mass matrices will exhibit tribimaximal mixing.
 
The most general superpotential of the singlet superfields invariant under 
$\Sigma(81)$ is given by
\begin{eqnarray}
W &=& m_\sigma (\sigma_1 \bar{\sigma}_1 + \sigma_2 \bar{\sigma}_2 + 
\sigma_3 \bar{\sigma}_3) + m_\chi (\chi_1 \bar{\chi}_1 + \chi_2 \bar{\chi}_2 
+ \chi_3 \bar{\chi}_3) \nonumber \\
&+& {1 \over 3} f (\sigma_1^3 + \sigma_2^3 + \sigma_3^3) + {1 \over 3} 
\bar{f} (\bar{\sigma}_1^3 + \bar{\sigma}_2^3 + \bar{\sigma}_3^3) 
+ {1 \over 3} h (\chi_1^3 + \chi_2^3 + \chi_3^3) + {1 \over 3} \bar{h} 
(\bar{\chi}_1^3 + \bar{\chi}_2^3 + \bar{\chi}_3^3) \nonumber \\ 
&+& \lambda (\chi_1 \sigma_2 \sigma_3 + \chi_2 \sigma_3 \sigma_1 + 
\chi_3 \sigma_1 \sigma_2) + \bar{\lambda} (\bar{\chi}_1 \bar{\sigma}_2 
\bar{\sigma}_3 + \bar{\chi}_2 \bar{\sigma}_3 \bar{\sigma}_1 + \bar{\chi}_3 
\bar{\sigma}_1 \bar{\sigma}_2). 
\end{eqnarray}
The resulting scalar potential has a supersymmetric minimum ($V=0$) if 
\begin{eqnarray}
0 &=& m_\sigma \bar{\sigma}_1 + f \sigma_1^2 + \lambda (\chi_2 \sigma_3 + 
\chi_3 \sigma_2) =  m_\sigma {\sigma}_1 + \bar{f} \bar{\sigma}_1^2 + 
\bar{\lambda} (\bar{\chi}_2 \bar{\sigma}_3 + \bar{\chi}_3 \bar{\sigma}_2), \\ 
0 &=& m_\sigma \bar{\sigma}_2 + f \sigma_2^2 + \lambda (\chi_3 \sigma_1 + 
\chi_1 \sigma_3) =  m_\sigma {\sigma}_2 + \bar{f} \bar{\sigma}_2^2 + 
\bar{\lambda} (\bar{\chi}_3 \bar{\sigma}_1 + \bar{\chi}_1 \bar{\sigma}_3), \\ 
0 &=& m_\sigma \bar{\sigma}_3 + f \sigma_3^2 + \lambda (\chi_1 \sigma_2 + 
\chi_2 \sigma_1) = m_\sigma {\sigma}_3 + \bar{f} \bar{\sigma}_3^2 + 
\bar{\lambda} (\bar{\chi}_1 \bar{\sigma}_2 + \bar{\chi}_2 \bar{\sigma}_1), \\ 
0 &=& m_\chi \bar{\chi}_1 + h \chi_1^2 + \lambda \sigma_2 \sigma_3 = 
m_\chi {\chi}_1 + \bar{h} \bar{\chi}_1^2 + \bar{\lambda} \bar{\sigma}_2 
\bar{\sigma}_3, \\ 
0 &=& m_\chi \bar{\chi}_2 + h \chi_2^2 + \lambda \sigma_3 \sigma_1 = 
m_\chi {\chi}_2 + \bar{h} \bar{\chi}_2^2 + \bar{\lambda} \bar{\sigma}_3 
\bar{\sigma}_1, \\ 
0 &=& m_\chi \bar{\chi}_3 + h \chi_3^2 + \lambda \sigma_1 \sigma_2 = 
m_\chi {\chi}_3 + \bar{h} \bar{\chi}_3^2 + \bar{\lambda} \bar{\sigma}_1 
\bar{\sigma}_2.
\end{eqnarray}
In the limit $\lambda=\bar{\lambda}=0$, the symmetry of $W$ is enlarged to 
$\Sigma(81) \times \Sigma(81)$.  Thus it is natural to expect $\lambda, 
\bar{\lambda} << f, \bar{f}, h, \bar{h}$, and as a first approximation, 
a possible solution of $V=0$ is 
\begin{eqnarray}
&& \langle \sigma_{1,2,3} \rangle_0 = -m_\sigma (f^2 \bar{f})^{-1/3}, ~~~ 
\langle \bar{\sigma}_{1,2,3} \rangle_0 = -m_\sigma (\bar{f}^2 {f})^{-1/3}, \\
&& \langle \chi_{1} \rangle_0 = -m_\chi (h^2 \bar{h})^{-1/3}, ~~~ 
\langle \bar{\chi}_{1} \rangle_0 = -m_\chi (\bar{h}^2 {h})^{-1/3}, ~~~ 
\langle \chi_{2,3} \rangle_0 = \langle \bar{\chi}_{2,3} \rangle_0 = 0,
\end{eqnarray}
where $\lambda = \bar{\lambda} = 0$ has been assumed.  [The appearance of 
domain walls can be avoided by explicit soft supersymmetry breaking terms 
which also break $\Sigma(81)$.]  This results in the 
desirable Yukawa terms $(l_1 + l_2 + l_3) l^c_1 \phi^0$, $(l_1 + 
\omega^2 l_2 + \omega l_3) l^c_2 \phi^0$, $(l_1 + \omega l_2 + \omega^2 l_3) 
l^c_3 \phi^0$, $(\nu_1 \nu_1 + \nu_2 \nu_2 + \nu_3 \nu_3) \xi^0$, and 
$\nu_2 \nu_3 \xi^0$, leading to tribimaximal mixing \cite{m04}.
Specifically
\begin{equation}
{\cal M}_l = \pmatrix{h_e & h_\mu & h_\tau \cr h_e & 
\omega^2 h_\mu& \omega h_\tau \cr h_e & \omega h_\mu & 
\omega^2 h_\tau} v = {1 \over \sqrt{3}} \pmatrix{1 & 1 & 1 \cr 1 & \omega^2 
& \omega \cr 1 & \omega & \omega^2} \pmatrix{h_e & 0 & 0 \cr 0 & h_\mu & 0 \cr 
0 & 0 & h_\tau} \sqrt{3} v,
\end{equation}
and \cite{af05,af06}
\begin{equation}
{\cal M}_\nu = \pmatrix{a & 0 & 0 \cr 0 & a & d \cr 0 & d & a} = 
\pmatrix{0 & 1 & 0 \cr 1/\sqrt{2} & 0 & i/\sqrt{2} \cr 1/\sqrt{2} & 0 & 
-i/\sqrt{2}} \pmatrix{a+d & 0 & 0 \cr 0 & a & 0 \cr 0 & 0 & -a+d} 
\pmatrix{0 & 1/\sqrt{2} & 1/\sqrt{2} \cr 1 & 0 & 0 \cr 0 & i/\sqrt{2} & 
-i/\sqrt{2}},
\end{equation}
thereby leading to Eq.~(1), i.e.
\begin{equation}
{1 \over \sqrt{3}} \pmatrix{1 & 1 & 1 \cr 1 & \omega & \omega^2 \cr 1 & 
\omega^2 & \omega} \pmatrix{0 & 1 & 0 \cr 1/\sqrt{2} & 0 & i/\sqrt{2} 
\cr 1/\sqrt{2} & 0 & -i/\sqrt{2}} = \pmatrix{ \sqrt{2/3} & \sqrt{1/3} & 
0 \cr -\sqrt{1/6} & \sqrt{1/3} & -\sqrt{1/2} \cr -\sqrt{1/6} & \sqrt{1/3} 
& \sqrt{1/2}}.
\end{equation}

This is thus another version of a successful derivation of tribimaximal 
mixing, but as in all previous such models, the predicted value of 
$\tan^2 \theta_{12} = 0.5$ is not the central value of present experimental 
data: $\tan^2 \theta_{12} = 0.45 \pm 0.05$.  To obtain a deviation from 
$\tan^2 \theta_{12} = 0.5$ in the context of $\Sigma(81)$ alone, consider 
now $\lambda, \bar{\lambda} \neq 0$ but small.  In that case,
\begin{equation}
{\langle \chi_{2,3} \rangle \over \langle \chi_1 \rangle_0} \simeq 
{\bar{\lambda} (h^2 \bar{h})^{1/3} \over (f^2 \bar{f})^{2/3}} {m_\sigma^2 
\over m_\chi^2}, ~~~ {\langle \bar{\chi}_{2,3} \rangle \over \langle 
\bar{\chi}_1 \rangle_0} \simeq {{\lambda} (\bar{h}^2 {h})^{1/3} \over 
(\bar{f}^2 {f})^{2/3}} {m_\sigma^2 \over m_\chi^2},
\end{equation}
\begin{equation}
\delta \langle \sigma_1 \rangle \simeq 0, ~~~ {\delta \langle \sigma_{2,3} 
\rangle \over \langle \sigma \rangle_0} \simeq -{m_\chi \over 3 m_\sigma} 
\left[ {\bar{\lambda} f^{1/3} \over (\bar{h}^2 h \bar{f})^{1/3}} + 
{2 \lambda \bar{f}^{1/3} \over (h^2 \bar{h} f)^{1/3}} \right],
\end{equation}
\begin{equation}
\delta \langle \bar{\sigma}_1 \rangle \simeq 0, ~~~ {\delta \langle 
\bar{\sigma}_{2,3} \rangle \over \langle \bar{\sigma} \rangle_0} \simeq 
-{m_\chi \over 3 m_\sigma} \left[ {{\lambda} \bar{f}^{1/3} \over 
({h}^2 \bar{h} {f})^{1/3}} + {2 \bar{\lambda} {f}^{1/3} 
\over (\bar{h}^2 {h} \bar{f})^{1/3}} \right].
\end{equation}

Since $\langle \bar{\sigma}_1 \rangle \neq \langle \bar{\sigma}_{2,3} 
\rangle$, the charged-lepton mass matrix is modified.  Instead of 
Eq.~(17), it is now of the form
\begin{equation}
{\cal M}_l = \pmatrix{h_e v_1 & h_\mu v_1 & h_\tau v_1 \cr h_e v_2 & 
\omega^2 h_\mu v_2 & \omega h_\tau v_2 \cr h_e v_2 & \omega h_\mu v_2 & 
\omega^2 h_\tau v_2}.
\end{equation}
Using the phenomenological hierarchy $h_e << h_\mu << h_\tau$, it is easily 
shown \cite{m06_4}, to first approximation, that the tribimaximal 
$U_{\alpha i}$ of Eq.~(1) is multiplied on the left by
\begin{equation}
R = \pmatrix{1 & -r & -r \cr r & 1 & -r \cr r & r & 1}, ~~~ r \simeq {v_1-v_2 
\over v_1+2v_2} \simeq - {\delta \langle \bar{\sigma}_{2,3} \rangle 
\over 3 \langle \bar{\sigma} \rangle_0}.
\end{equation}
The neutrino mass matrix of Eq.~(18) is 
also changed, i.e.
\begin{equation}
{\cal M}_\nu = \pmatrix{a & e & e \cr e & a+b & d \cr e & d & a+b},
\end{equation}
where $|b| << |a|$ and $|e| << |d|$.  This leads to a correction of Eq.~(1) 
on the right by the matrix
\begin{equation}
R' = \pmatrix{1 & -r' & 0 \cr r' & 1 & 0 \cr 0 & 0 & 1}, ~~~ r' \simeq 
{\sqrt{2} e \over d} \simeq {\sqrt{2} \langle \chi_{2,3} \rangle \over 
\langle \chi_1 \rangle_0}.
\end{equation}
Hence the corrected mixing matrix is 
given by
\begin{equation}
U_{\alpha i} \simeq \pmatrix{ \sqrt{2/3}(1+r+r'/\sqrt{2}) & \sqrt{1/3}(1-2r-
\sqrt{2}r') & 0 \cr -\sqrt{1/6}(1-3r-\sqrt{2}r') & \sqrt{1/3}(1+r'/\sqrt{2}) 
& -\sqrt{1/2}(1+r) \cr -\sqrt{1/6}(1-r-\sqrt{2}r') & \sqrt{1/3}(1+2r+
r'/\sqrt{2}) & \sqrt{1/2}(1-r)}.
\end{equation}
Therefore,
\begin{equation}
\tan^2 \theta_{12} \simeq {1 \over 2} - 3(r+r'/\sqrt{2}), ~~~ 
\tan^2 \theta_{23} \simeq 1+4r, ~~~ \theta_{13} \simeq 0.
\end{equation}
For example, let $r = r' = 0.01$, then $\tan^2 \theta_{12} \simeq 0.45$, 
whereas $\tan^2 \theta_{23} \simeq 1.04$ which is equivalent to 
$\sin^2 2 \theta_{23} \simeq 0.9996$.  A better match to the data is thus 
obtained.

In summary, it has been shown in this paper that neutrino tribimaximal 
mixing is a natural limit in a supersymmetric model based on $\Sigma(81)$ 
without any imposed auxiliary symmetry as in previous models.  The 
assumed particle content and the requirement of supersymmetry allow a 
prediction (tribimaximal mixing) in the limit that the superpotential 
has the enlarged symmetry $\Sigma(81) \times \Sigma(81)$.  The parameters 
which break $\Sigma(81) \times \Sigma(81)$ but preserve $\Sigma(81)$ are 
assumed to be small to check the possible deviation from tribimaximal 
mixing.  The conclusions are that whereas corrections are expected, they 
are such that only $\tan^2 \theta_{12}$ may deviate significantly from 0.5 
(say to 0.45) without affecting much the predictions $\sin^2 2 \theta_{23} 
= 1$ and $\theta_{13} = 0$.  [It should be pointed out that in many 
models, the {\it ad hoc} assumption $\nu_2 = (\nu_e + \nu_\mu + \nu_\tau)
/\sqrt{3}$ is made. In that case, $\tan^2 \theta_{12}$ must be greater 
than 0.5, not smaller.]

This work was supported in part by the U.~S.~Department of Energy under Grant 
No.~DE-FG03-94ER40837.

\noindent {\bf Appendix}

The 9 one-dimensional irreducible representations together with ${\bf 3_D}$, 
${\bf \bar{3}_D}$ behave as in $\Delta(27)$, i.e. \cite{m06_6}
\begin{equation}
{\bf 3_D} \times {\bf 3_D} = {\bf \bar{3}_D} + {\bf \bar{3}_D} + 
{\bf \bar{3}_D}, ~~~  {\bf 3_D} \times {\bf \bar{3}_D} = {\bf 1_{1,2,3}} 
+ {\bf 1_{4,5,6}} + {\bf 1_{7,8,9}}.
\end{equation}
The ${\bf 3_A},{\bf 3_B},{\bf 3_C}$ representations are cyclically 
equivalent, as are their conjugates. Their multiplication rules are
\begin{eqnarray}
&& {\bf 3_A} \times {\bf \bar{3}_A} = {\bf 3_B} \times {\bf \bar{3}_B} = 
{\bf 3_C} \times {\bf \bar{3}_C} = {\bf 1_{1,2,3}} + {\bf 3_D} + 
{\bf \bar{3}_D}, \\
&& {\bf 3_B} \times {\bf \bar{3}_A} = {\bf 3_C} \times {\bf \bar{3}_B} = 
{\bf 3_A} \times {\bf \bar{3}_C} = {\bf 1_{4,5,6}} + {\bf 3_D} + 
{\bf \bar{3}_D}, \\
&& {\bf 3_C} \times {\bf \bar{3}_A} = {\bf 3_A} \times {\bf \bar{3}_B} = 
{\bf 3_B} \times {\bf \bar{3}_C} = {\bf 1_{7,8,9}} + {\bf 3_D} + 
{\bf \bar{3}_D}, \\ 
&& {\bf 3_A} \times {\bf {3}_A} = {\bf 3_B} \times {\bf {3}_C} = 
{\bf \bar{3}_A} + {\bf \bar{3}_B} + {\bf \bar{3}_B}, \\
&& {\bf 3_B} \times {\bf {3}_B} = {\bf 3_C} \times {\bf {3}_A} = 
{\bf \bar{3}_B} + {\bf \bar{3}_C} + {\bf \bar{3}_C}, \\
&& {\bf 3_C} \times {\bf {3}_C} =  {\bf 3_A} \times {\bf {3}_B} = 
{\bf \bar{3}_C} + {\bf \bar{3}_A} + {\bf \bar{3}_A}.
\end{eqnarray}

\bibliographystyle{unsrt}

\end{document}